\documentclass[prapplied,rsi, amsmath,amssymb,reprint,superscriptaddress]{revtex4-2}

\usepackage{longtable}
\usepackage{graphicx}
\usepackage{color}
\usepackage{soul}

\usepackage[version=3]{mhchem} 
\usepackage{blindtext}
\usepackage{bm}
\usepackage{array}
\newcolumntype{C}{>{\centering\arraybackslash}p{4em}}
\newcolumntype{D}{>{\centering\arraybackslash}p{6em}}

\begin{document}
\title{Competing nucleation pathways in nanocrystal formation} 

\author{Carlos R. Salazar}
\affiliation{Univ. Lille, CNRS, INRA, ENSCL, UMR 8207, UMET, Unité Matériaux et Transformations, F 59000 Lille, France}
\affiliation{Centre d’élaboration des Matériaux et d’Etudes Structurales, CNRS (UPR 8011), 29 rue Jeanne Marvig, 31055 Toulouse Cedex 4, France}

\author{Akshay Krishna Ammothum Kandy}
\affiliation{Centre d’élaboration des Matériaux et d’Etudes Structurales, CNRS (UPR 8011), 29 rue Jeanne Marvig, 31055 Toulouse Cedex 4, France}

\author{Jean Furstoss}
\affiliation{Univ. Lille, CNRS, INRA, ENSCL, UMR 8207, UMET, Unité Matériaux et Transformations, F 59000 Lille, France}

\author{Quentin Gromoff}
\affiliation{Centre d’élaboration des Matériaux et d’Etudes Structurales, CNRS (UPR 8011), 29 rue Jeanne Marvig, 31055 Toulouse Cedex 4, France}

\author{Jacek Goniakowski}
\affiliation{CNRS, Sorbonne Université, Institut des NanoSciences de Paris, UMR 7588, 4 Place Jussieu, F-75005 Paris, France}

\author{Julien Lam}
\email{julien.lam@cnrs.fr}	
\affiliation{Univ. Lille, CNRS, INRA, ENSCL, UMR 8207, UMET, Unité Matériaux et Transformations, F 59000 Lille, France}
\affiliation{Centre d’élaboration des Matériaux et d’Etudes Structurales, CNRS (UPR 8011), 29 rue Jeanne Marvig, 31055 Toulouse Cedex 4, France}

\newcommand{\JL}[1]{\textcolor{red}{{\,[\textbf JL:} #1]\,}}
\newcommand{\CS}[1]{\textcolor{blue}{{\,[\textbf CS:} #1]\,}}

\begin{abstract}

Despite numerous efforts from numerical approaches to complement experimental measurements, several fundamental challenges have still hindered one's ability to truly provide an atomistic picture of the nucleation process in nanocrystals. Among them, our study resolves three obstacles: (1)\,Machine-learning force fields including long-range interactions able to capture the finesse of the underlying atomic interactions, (2) Data-driven characterization of the local ordering in a complex structural landscape associated with several crystal polymorphs and (3) Comparing results from a large range of temperatures using both brute-force and rare-event sampling. Altogether, our simulation strategy has allowed us to study zinc oxide crystallization from nano-droplet melt. Remarkably, our results show that different nucleation pathways compete depending on the investigated degree of supercooling. 
\end{abstract}

\maketitle

\section{Introduction}

Polymorphisms occur when the same material can be found in different structural forms. In protein crystals, the competition between each of the possible structures has dramatic consequences causing amyloid diseases\,\cite{Petkova2005Jan,Close2018Feb,Fandrich2018Mar} and toxicity of pharmaceutical compounds\,\cite{Zhang2014Nov,Morissette2003Mar}. Meanwhile, for technological applications associated with material science, each crystal phase has distinct physical and chemical properties, necessitating the stabilization of a specific polymorphic form. As the triggering mechanisms for the emergence of order, crystal nucleation should have been key for controlling polymorphic selection. However, its study remains extremely challenging because disparate lengths and time scales are simultaneously involved\,\cite{Sosso2016Jun,Finney2023Nov}. On the one hand, in terms of size, an extended mother phase along with a small critical cluster made of few tens of atoms must be jointly studied. On the other hand, crystal nucleation combines both a long time scale for stochastic fluctuations to trigger the critical event and a short time scale for crystal growth.

This already complex picture is exacerbated for nanoscale systems. Indeed, the preponderance of surface effects expands the structural landscape of possible polymorphic structures. In addition, a competition between two different natures of nucleation can be found: homogeneous in the core and heterogeneous in the peripheral. While numerous experimental works have provided insights into this challenging nucleation process\,\cite{Ramamoorthy2020Aug,Schiener2015Jun,Ibrahimkutty2012Sep,Albrecht2021Mar}, numerical simulations should have been the ideal tool as it provides a dynamic picture at the single-particle level. However, simulating crystal nucleation in nanoparticles requires facing two major challenges. On the one hand, ab initio molecular dynamics simulations are too much computationally demanding to perform the necessary large-scale simulations while classical interaction potentials can not always precisely model both bulk and surface effects. On the other hand, nucleation involves overcoming a free energy barrier and is thus an intrinsically rare-event combining a long induction time (set by the nucleation rate) with a short transition period (set by the growth rate).

Our study focuses on the polymorphic competition in zinc oxide nanoparticles which exhibits promising electrochemical\,\cite{Zhou2023Apr,Nagajyothi2013Oct,Sun2016Nov} and antibacterial\,\cite{Matinise2017Jun,Pushpalatha2022,Islam2022Mar,Gudkov2021Mar} activities. For all these applications, a key feature of ZnO is its structural complexity associated with the downsizing to the nanoscale\,\cite{Wang2023Jun,Chen2018Aug,Zagorac2022Jan,Leitner2018Mar,Vines2017Jul}. Indeed, while the Wurtzite structure is the most stable in bulk, it was found that a competition between different polymorphs exists with body-centered tetragonal structure being more preponderant at sufficiently small sizes. However, body-centered tetragonal is yet to be observed experimentally in a nanoparticle. This crystal phase was first theoretically discovered by \citet{Wang2007Nov}, and it has been experimentally observed only on surface reconstructions\,\cite{He2012Jul} and nanosheets\,\cite{Wang2016Jan}. From the theoretical perspective, studies of the formation process of ZnO nanoparticles have mostly employed classical force-fields including ReaxFF and Buckingham\,\cite{Gao2023Feb, Baguer2009Aug, Barcaro2019Feb}. More recently, more precise modeling was also achieved using machine-learning interaction potentials\,\cite{Artrith2011Apr,Goniakowski2022Oct} (MLIP) yet never focusing on crystal nucleation at the nanoscale. 

In this work, we first constructed a machine-learning interaction potential including long-range interactions using the recently developed Physical LassoLars Interaction Potential (PLIP) methodology and demonstrated its higher accuracy when compared to simpler short-range MLIP. Then, we performed both brute-force molecular dynamics along with seeded simulations to unravel the competition between Wurtzite (WRZ) which is the most stable crystal in bulk, and body-centered tetragonal phase (BCT) which is a concurring phase only stable at sufficiently small nanoparticle sizes. Finally, to analyze the obtained results, we developed a data-driven clustering method based on a Gaussian-mixture model enabling for characterizing the local structure at the atomistic level. Altogether, by complementing brute-force simulations with the seeding approach, we managed to demonstrate the presence of two different nucleation pathways depending on the investigated temperatures: Multi-step process involving a metastable crystal phase and Classical nucleation picture respectively at high and moderate degrees of supercooling.

\section{Results}
\subsection{Validation of the machine-learning interaction potential}

Many variations of MLIPs have demonstrated excellent performance in capturing short-range interactions by utilizing descriptors of local atomic environments during their model training. Nevertheless, employing these short-ranged local environment descriptors in MLIPs can pose difficulties when trying to simulate systems that entail substantial long-range interactions\,\cite{Behler2021Jul,Behler2021Aug}. For this study of ZnO, we develop a PLIP+Q model that combines the PLIP approach for short-range interactions\,\cite{Kandy2023May,Goniakowski2022Oct,Benoit2020Dec,Tallec2023Jan} along with a scaled point charge model to incorporate a long-range description of the interactions. Please see Methods A for further details on the PLIP and PLIP+Q models. We note that the superiority of the short-range PLIP  against other classical force fields for ZnO was already determined in our previous work\,\cite{Goniakowski2022Oct}. 

\begin{figure}[ht]
\includegraphics[width=8.6cm]{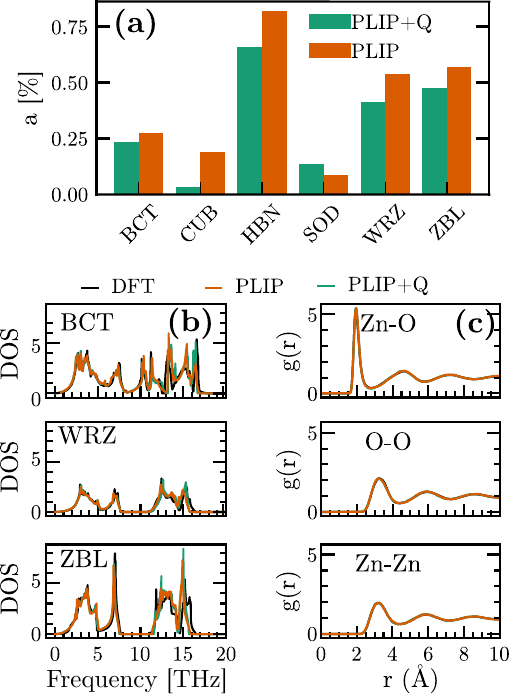}
\caption{Properties of bulk ZnO modeled with PLIP and PLIP+Q: (a) Lattice parameter, (b) Phonon density of states, and (c) Liquid radial distribution functions obtained at $1500$\,K. Please see Table\,\ref{CrystalStructure} for the nomenclature of each crystal phases.}
\label{fig:bulk}
\end{figure}

To begin, the error in lattice parameters with respect to density functional theory (DFT) calculations for all the ZnO polymorphs in the database are shown in Fig.\,\ref{fig:bulk}.a. In general, both PLIP+Q and PLIP method performs well, as the errors remain below 1 $\%$. Nevertheless,  PLIP+Q seems to slightly improve the lattice parameter compared to PLIP, except for the sodalite (SOD) polymorph. 

Further, we compute the phonon density of states (DOS) for WRZ, zincblend (ZBL), and BCT polymorphs using a supercell approach where atomic positions are slightly perturbed to measure the reaction forces. The following  super cells were used:  WRZ (5 × 5 × 3), ZBL(3 × 3 × 3), and BCT (3 × 3 × 5).  The phonon calculations are carried out using the PHONOPY package\,\cite{Togo2022Dec,Togo2023Jun}. Fig.\,\ref{fig:bulk}b, shows the comparison between PLIP, PLIP+Q, and DFT phonon density of states. Qualitatively, the DOS from both PLIP and PLIP+Q methods show a good match with the DFT reference. In particular, the low-frequency acoustic DOS is almost an exact match between PLIP and PLIP+Q. However, for high-frequency optical modes, PLIP+Q shows a better agreement than PLIP, for peaks between 16-17 Hz for BCT, 15-16 Hz for WRZ, and 12 Hz for ZBL. 

After studying the properties of ZnO crystals, we measure the accuracy of the model for disordered bulk structures. Ab initio molecular dynamics (AIMD) as well as MD  with the obtained MLIP are carried out averaging through three different initial structures in the NVT ensemble at $1500$\,K for 4\,ps and with a time-step of 1\,fs. The partial radial distribution functions (RDF) obtained from both PLIP and PLIP+Q show a good agreement with AIMD results [See Fig.\,\ref{fig:bulk}c]. It seems that a good description of short-range interactions can already provide structural information for disordered structures and that the addition of long-range interactions does not alter the agreement.

\begin{figure*}[ht]
\includegraphics[width=17cm]{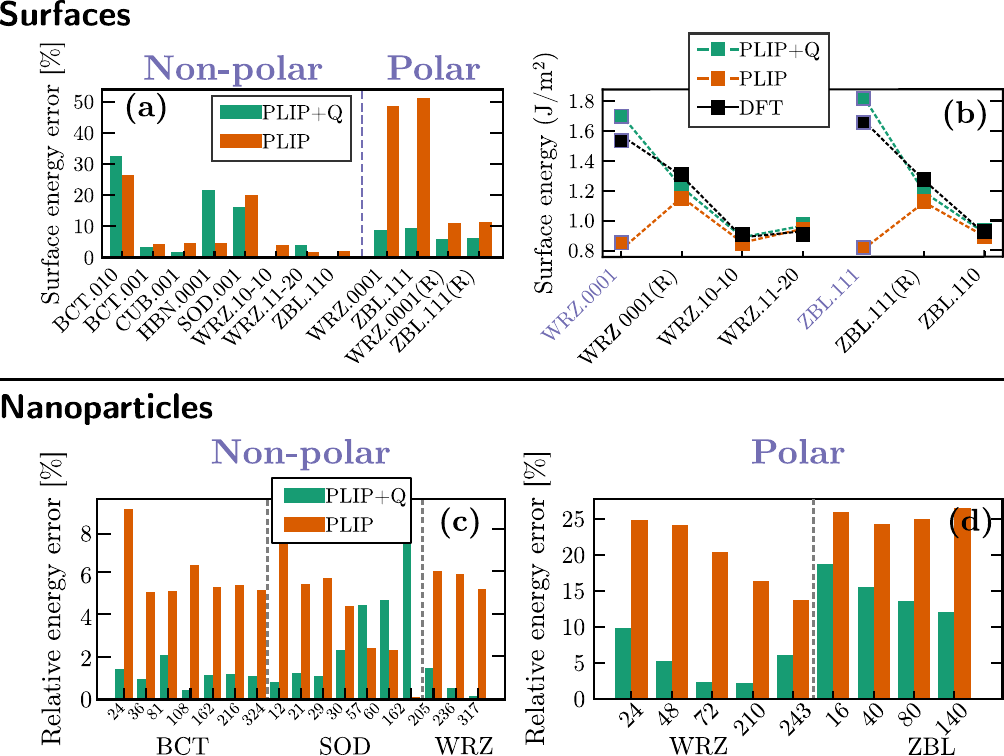}
\caption{(a,b) Solid-vacuum surface energy measured with PLIP and PLIP+Q for different ZnO polymorphs. (a) Surface energy error with non-polar and polar surfaces respectively on the left and on the right. (b) Value of the surface energy when focusing on WRZ and ZBL polymorphs. In violet, the polar surfaces are highlighted. "(R)" designates polar-reconstructed surfaces. (c,d)\,Nanoparticle energy when comparing PLIP and PLIP+Q. (c) Optimized nanoparticles without any polar surfaces as obtained by \citet{Vines2017Jul} and (d) Non-optimized nanoparticles created to exhibit polar surfaces.}
\label{fig:PLIP2}
\end{figure*}

Thus far, the performance of both PLIP and PLIP+Q has been almost indistinguishable. However, in the context of nucleation of nanoparticles, the relative solid-vacuum surface energies of low-index surfaces dictate the obtained morphology and should therefore be well reproduced too. In the specific case of zinc oxide, one must put an emphasis on studying non-polar as well as low-index polar and polar-reconstructed surfaces. On the left part of Fig.\,\ref{fig:PLIP2}.a, we show results for non-polar surfaces where one can see that the behavior of PLIP and PLIP+Q closely aligns. Then, for polar-reconstructed surfaces, although PLIP is able to perform quite well, PLIP+Q still provides slightly better agreement with the DFT results. However, a noticeable distinction emerges for polar surfaces, where PLIP+Q largely outperforms PLIP. Specifically, PLIP exhibits an error larger than 50 percent for those non-reconstructed polar surfaces, while the error is less than 10 percent for the PLIP+Q. More importantly, PLIP not only exhibits large percentage errors, it also incorrectly predicts that the two studied polar surfaces are the most stable ones [See Fig.\,\ref{fig:PLIP2}b]. In contrast, PLIP+Q is able to retrieve the correct stability ordering when compared to DFT calculations, as demonstrated in Fig.\,\ref{fig:PLIP2}.b.

Prompted by these results, we conduct an assessment of the performance of both PLIP and PLIP+Q on nanostructures. To begin, we use ZnO clusters obtained by \citet{Vines2017Jul}.  We explore 3 different families of (ZnO)$_N$ structures, encompassing cuts of bulk crystalline structures namely BCT, WRZ, and SOD. The nanoparticles are optimized further with our DFT setting. The error in energy is measured for each system by comparing to the DFT optimized reference and MLIP optimized structure and is shown in Fig.\,\ref{fig:PLIP2}.c. Although the training set is composed of bulk and surface configurations only, it is evident from Fig.\,\ref{fig:PLIP2}.c that both MLIPs are transferable to these nanometric structures. Since these nanoparticles do not exhibit any polar surfaces, we additionally test both MLIPs using WRZ and ZBL nanostructures purposely constructed to expose polar surface terminations. In the case of ZBL, we design octahedral nanoparticles with polar (111) facets, featuring truncated corners to ensure their overall stoichiometry. The WRZ nanoparticles are made by top-down cuts of the bulk polymorph, as illustrated by \citet{Vines2017Jul}. From Fig.\ref{fig:PLIP2}.d, which displays the corresponding single point energy errors, it can be seen that while PLIP was able to correctly model nanoparticles with non-polar surfaces, it leads to much higher error than PLIP+Q for both types of nanoparticles exhibiting polar surfaces. The discrepancy is mainly driven by the substantial underestimation of surface energy of polar terminations in the local PLIP approach, and may lead to a spurious abundance of polar nanoparticles in MD simulations.

%

%
%


Altogether, the main drawback of PLIP is that polar surfaces are not only incorrectly reproduced in energy, but they are also considered the most stable ones. Such an issue has dramatic consequences when dealing with nanostructures. However, although very simple in its conceptual formalism, PLIP+Q is already able to rectify this error and model correctly both polar and non-polar surfaces as well as their subsequent nanostructures. Consequently, moving forward, the results presented in the remainder of the article will exclusively focus on calculations obtained using the PLIP+Q potential. We note that the computational efficacy was reduced by roughly 20\,\% upon adding the long-range interactions.

\section{Brute-force simulations}

\begin{figure*}[ht]
    \centering
    \includegraphics[width=17cm]{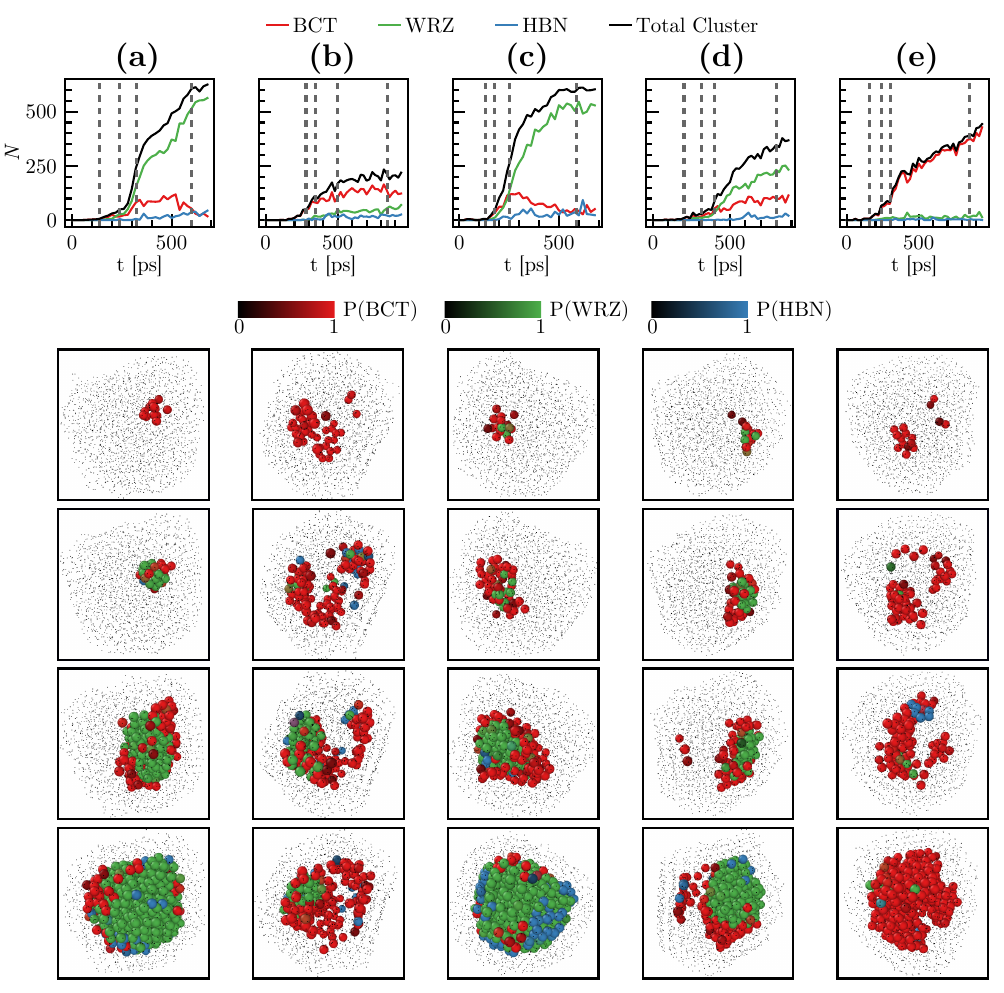}
    \caption{Brute force nucleation simulations of ZnO nanoparticles made of 2000 atoms. Five simulations are shown using different random instances for the initial velocities in graphs (a-e) where the simulation time is shown on the x-axis and the number of atoms is shown on the y-axis. Snapshots of the clusters are shown below each graph corresponding to times indicated by the vertical dashed lines. Atoms are colored according to their probabilities to be in BCT structure (red), WRZ structure (green), and h-BN (HBN) structure (blue). Images are plotted using Ovito software\,\cite{Stukowski2009Dec}.}
    \label{fig:bruteForce}
\end{figure*}

To study the crystallization of ZnO nanoparticles we perform brute-force simulations for liquid nano-droplets of $500$, $1000$, $1500$, and $2000$ atoms at the same degree of supercooling ie. $T/T_{melt}=0.625$. The simulations are carried out using different random instance for the initial atomic velocities so that different nucleation pathways could be explored. Please see Methods B for further details on the brute-force simulation protocol. Then, in order to analyze the structural composition of the system during nucleation and growth, we develop a supervised machine-learning method based on the combination of Gaussian mixture model for classification and Steinhardt bond order parameters for structural description. Further details on the procedure are to be found in Methods C. The composition of the nanoparticles as a function of time is shown for the system of $2000$ atoms in Fig.\,\ref{fig:bruteForce} as well as snapshots at different key points in time. In all the obtained simulations, one can observe an induction time of approximately $300$\,ps after which a first nucleus sufficiently large is formed. Surprisingly, while WRZ is the most stable crystal structure both in bulk and in this size regime, the nuclei consist primarily [See Fig.\,\ref{fig:bruteForce}(c,d)], and in some cases completely [See Fig.\,\ref{fig:bruteForce}(a,b,e)], of atoms in the BCT structure. Additionally, in three cases [See Fig.\,\ref{fig:bruteForce}(a,c,e)] where only one nuclei emerges at the early stages, WRZ competes and becomes the most preponderant crystal phase at the later stages. Meanwhile, in two others cases [See Fig.\,\ref{fig:bruteForce}(b,d)] where more than one nuclei are formed, the system presents a slower growth rate and during the entire simulation is mostly composed of atoms in the BCT structure. These results clearly show that there is a competition in the formation of the BCT and WRZ crystal phases, where BCT is more predominant in the early stages of crystallization while WRZ forms later and becomes the main structure.

These observations are further supported by additional brute-force simulations of $500$, $1000$, and $1500$ atoms [See Supplementary Figures (1-3)]. In all systems sizes, BCT forms first and WRZ appears later in some of the simulations. Such a two-step nucleation process, observed here in the crystal nucleation from a liquid nano-droplet, is consistent with our previous results obtained in bulk and with short-range PLIP\,\cite{Goniakowski2022Oct}. It is also reminiscent of seminal findings in much more simple systems interacting with Lennard-Jones and hard-spheres force fields where, while the face-centered cubic is the most thermodynamically stable, it is the body-centered cubic that is often observed at the early stages of nucleation\,\cite{Pusey2009Dec,Sanz2011May,Trudu2006Sep,Desgranges2007Jun}.

\section{Seeding simulations}

In the brute-force simulations, simulations must be performed at very low temperatures ie. in a deeply supercooled regime in order thus reducing both the free energy barrier and the associated induction time. In order to further investigate this competition and to verify its presence at moderate degrees of supercooling, one has to employ rare-event sampling techniques and we therefore perform seeding simulations\,\cite{ref:espinosa2}. In particular, a crystal seed is manually inserted in the fluid and its critical temperature is characterized as the temperature for which the seed neither shrinks nor grows\,\cite{ref:espinosa1}. By inserting a crystal seed, the free energy barrier of nucleation is artificially overcome, allowing for the study of more realistic conditions closer to the melting temperature. The seeding technique has already been applied to the study of crystal nucleation in many different systems ranging from condensed to soft matters\,\cite{ref:BaiLi2, ref:seeding1, ref:seeding2, ref:seeding3, ref:seeding4, ref:espinosa1, ref:seeding5}. But, to the best of our knowledge, the seeding technique has never been applied to nanocrystal simulations. Herein, we used the seeding technique to find the critical temperature of WRZ and BCT crystalline clusters which will allow us to address the competition between these two crystalline structures in nanoparticles [Please see Methods B for more details on the seeding simulation protocol]. 

\begin{figure*}[ht]
    \centering
    \includegraphics[width=17cm]{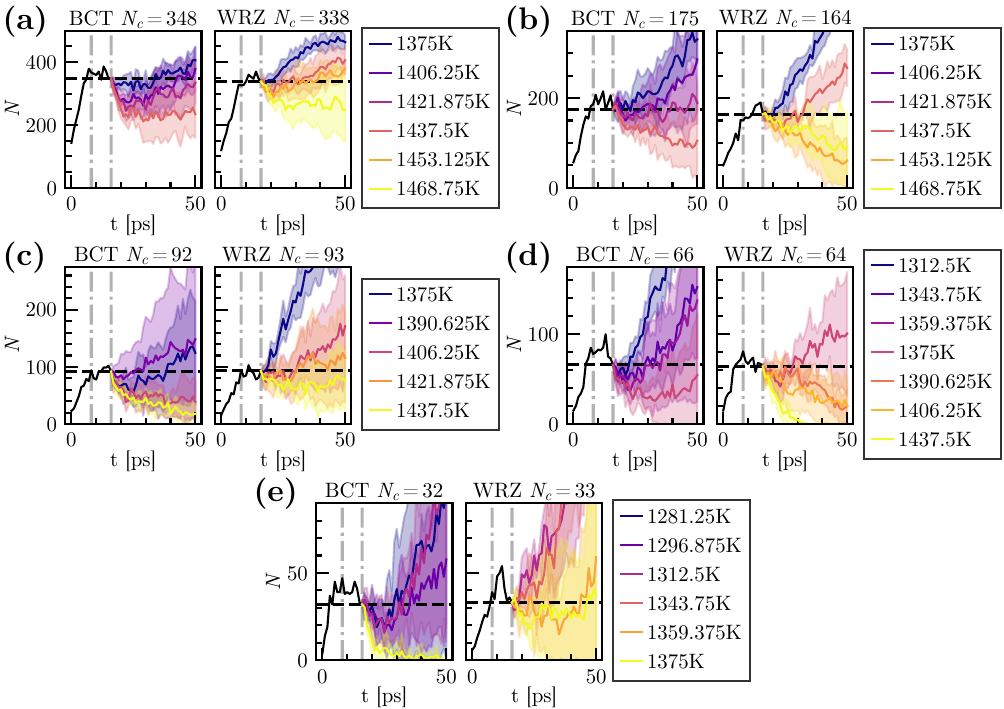}
    \caption{(a-e) Growth/Melt simulations for BCT and WRZ clusters of different sizes are shown in subfigures with a droplet made of 2000 atoms. The relaxation steps are shown in each graph with a black line, after which the growth/melt simulations are shown for different temperatures. In each subfigure, two clusters of similar critical cluster size $N_c$ but different crystal structures are compared. In particular, on the left (resp. on the right), the seed is made of BCT (resp. WRZ) crystalline atoms. Meanwhile, for sub-figures a to e, size of the crystalline seeds is respectively around 343, 170, 92, 65 and 32 atoms.}
    \label{fig:growth}
\end{figure*}

Results of the growth/melting curves are shown in Fig.\,\ref{fig:growth} for a 2000-atoms system and for initial crystal seeds of different sizes and crystal structures [Please see Supplementary Figures (4-6) for results obtained with different numbers of atoms inside the droplet]. For each growth/melting curves, standard deviations are showed in shaded colors and are obtained through 5 different simulations using the same state conditions (initial crystalline seed, temperature and total droplet size) yet with different initial velocities. First, we note that in all seeding simulations, the size of the biggest crystalline cluster $N$ increases during the relaxation step. This is required to fine-tune the initial size $N_0$ in order to reach the desired studied size $N_{c}$. Then, despite this fine-tuning of the relaxation process, Fig.\,\ref{fig:growth} shows that for a given set of conditions (i.e. $N_{c}$, $T$ and crystal structure), there is still a large standard deviation around each plain line and subsequently a large discrepancy between simulations with different initial velocity conditions. This justifies the necessity to generate several growth/melting curves per set of conditions ($5$ in our case). 

\begin{figure*}[ht]
    \centering
    \includegraphics[width=17cm]{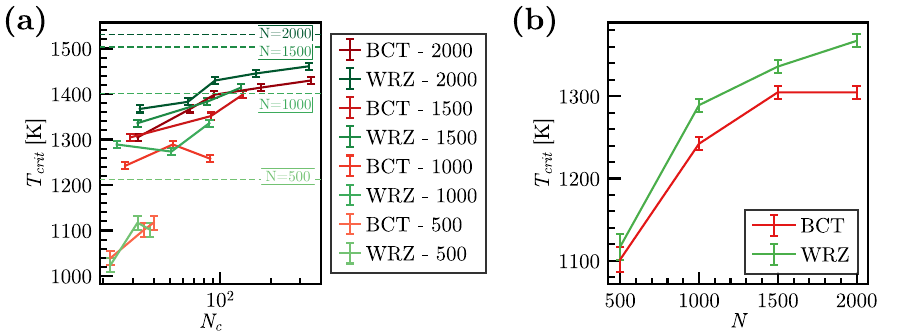}
    \caption{(a) Critical temperature results for different cluster structures and droplet sizes as a function of critical cluster size. The critical temperature is shown on the y-axis while the critical cluster size is shown in logarithmic scale on the x-axis. The melting temperatures of WRZ nanoparticles are shown in dashed lines where the color represents the size of the nanoparticle. (b) Dependence of the critical temperature on the nano-droplet size using a similar initial crystal seed size of approximately 30 atoms.}
    \label{fig:critTemps}
\end{figure*}

For each of the inserted seeds, one can deduce from the growth/melting curves a critical temperature $T_{crit}$ located between the lowest temperature at which the cluster shrinks and the highest temperature at which it grows. Ideally, the critical temperature is such that the slope of $N$ as a function of time is zero. In practice, a first temperature range is roughly selected and growth/melting simulations are performed at the upper and lower limits of the range, as well as in the middle. Then, the cluster sizes are plotted and the temperature for the next simulation is chosen as the value in between the temperatures with opposite slopes of $N$. We iterate this method until reaching a temperature range of $16$\,K which defines our reported error on the measurement of $T_{crit}$ and we use the middle between the two extreme values of the obtained range. The critical temperatures found for different crystalline cluster sizes are shown in Fig.\,\ref{fig:critTemps} together with the critical temperatures found for seeds in differently sized droplets. Using the results for the 2000-atoms system and our estimate of the melting temperature, the system is located in a degree of supercooling $T/T_{melt}$ between 0.8 and 0.9 which is way larger compared to the 0.625 studied with the brute-force simulations. The first observation is that the critical temperature increases with size meaning that lower temperatures are necessary to stabilize the smallest critical seeds. For all cluster sizes in the 2000-atoms and 1500-atoms systems, WRZ has a significantly higher critical temperature when compared to BCT. Accordingly, there is a range of temperatures where a crystal seed of a similar size will be more stable in the WRZ phase than in the BCT phase. In the case of the 1000-atoms and 500-atoms systems, we observe that the difference in critical temperature between BCT and WRZ crystal clusters becomes less significant. More specifically, it is observed that the critical temperatures for BCT and WRZ on the 500-atoms system converge, which is consistent with what has been observed in the brute-force simulations and energy minimization at $0$\,K\,\cite{Vines2017Jul}. In addition, the largest BCT crystal seed made of approximately 300 atoms in the 2000-atoms system shows an unstable behavior by sharply decreasing in size at the beginning of the growth/melting simulation [See Fig.\,\ref{fig:growth}.a]. As such, it can be conjectured that at these higher temperature regimes where the free energy barrier is the highest, the only possible nucleation pathway is the one starting with WRZ seeds. In closing, Fig.\,\ref{fig:critTemps}.b shows that a strong dependence of the critical temperature on the system size as for a similar critical cluster size there is a difference of almost $300$\,K in critical temperature between the 500-atoms and 2000-atoms system. This finding strongly suggest that the nanoscale reduction associated with finite-size effects and surface preponderance is already at play in the investigated size regime.


Ultimately, our seeding simulations reveal a completely different nucleation behavior compared to brute force results that focused in a more deeply supercooled regime. Indeed, we showed that BCT and WRZ are respectively favored in deeply supercooling conditions (brute-force simulations) and in moderate supercooling conditions (seeding simulations). Therefore,  nucleation mechanisms are highly driven by the investigated degree of supercooling thus advocating for the necessity to combine brute force and rare-event sampling approaches. A similar observation was also made when studying nucleation from a dilute phase as in a gas or in the presence of a non-reactive solvent like NaCl in water. In both cases, depending on the saturation regime, one can indeed either directly nucleate crystalline clusters or start with a so-called high-density amorphous precursor\,\cite{Bulutoglu2022May,Addula2021Apr,Jiang2019Mar,Iida2023Apr}.

\section{Growth mechanisms}

Along with determining the critical temperature as a function of the critical size, seeding simulations also allow for studying the subsequent growth mechanisms. Herein, we extended the previous growth curves during $84$\,ps after the relaxation procedure and focused on temperature regimes where the crystal seed grows. 

\begin{figure*}[ht]
    \centering
    \includegraphics[width=17cm]{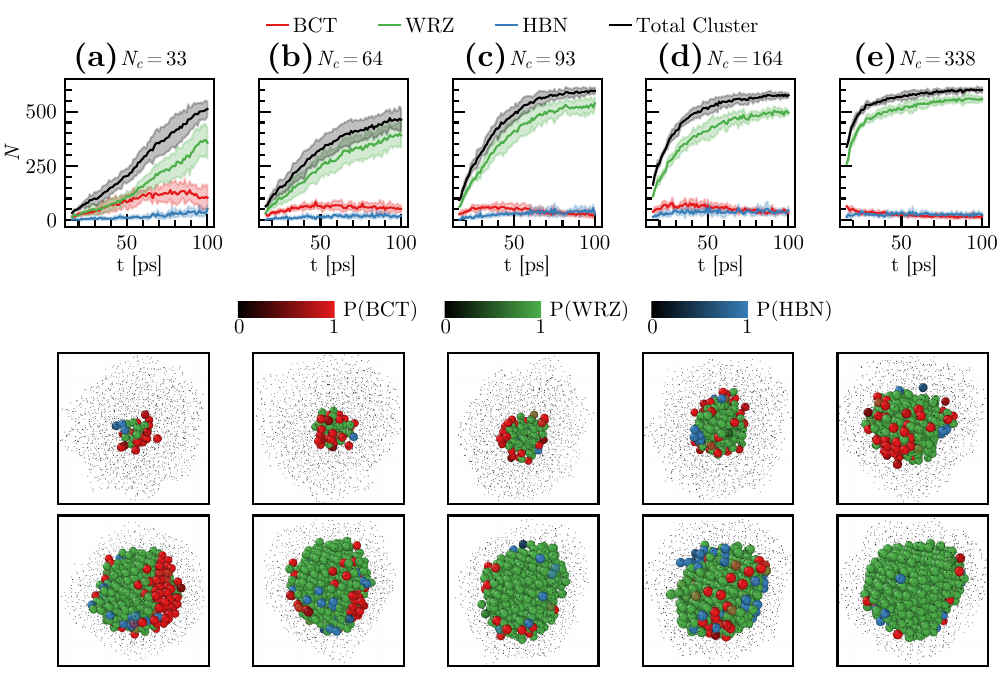}
    \caption{Extended growth simulations following the insertion of a WRZ crystal seed at $1000$\,K for five different seed sizes in a 2000-atoms system. The composition of the crystalline clusters is shown in the graphs in each column. The first and last snapshots of each simulation are shown below the corresponding graphs. Atoms are colored according to their probabilities to be in BCT structure (red), WRZ structure (green), and HBN structure (blue).}
    \label{fig:growthwrz}
\end{figure*}

At first, in the WRZ case, Fig.\,\ref{fig:growthwrz} shows that the cluster composition was consistent for all of the considered seed sizes ie. the cluster core is composed of atoms in the WRZ phase at the beginning and remains so until the end of the simulations. This is shown on the different graphs on which it can be noted that the number of atoms in the WRZ phase is always equal to or bigger than the number of atoms in the BCT phase. On the other hand, the outer layers of the cluster are often composed of atoms in the BCT and h-BN (HBN) phases. These atoms are located at the interface between the crystal and the liquid. The presence of few atoms predicted to be in the HBN phase at the surface of the crystalline cluster is probably a consequence of the large resemblance between HBN and WRZ structures. On the other hand, the scarce presence of atoms in the BCT phase is reminiscent of the BCT vs. WRZ competition. We note that these BCT atoms are mostly located at the surface of the crystalline cluster. 

\begin{figure*}[ht]
    \centering
    \includegraphics[width=17cm]{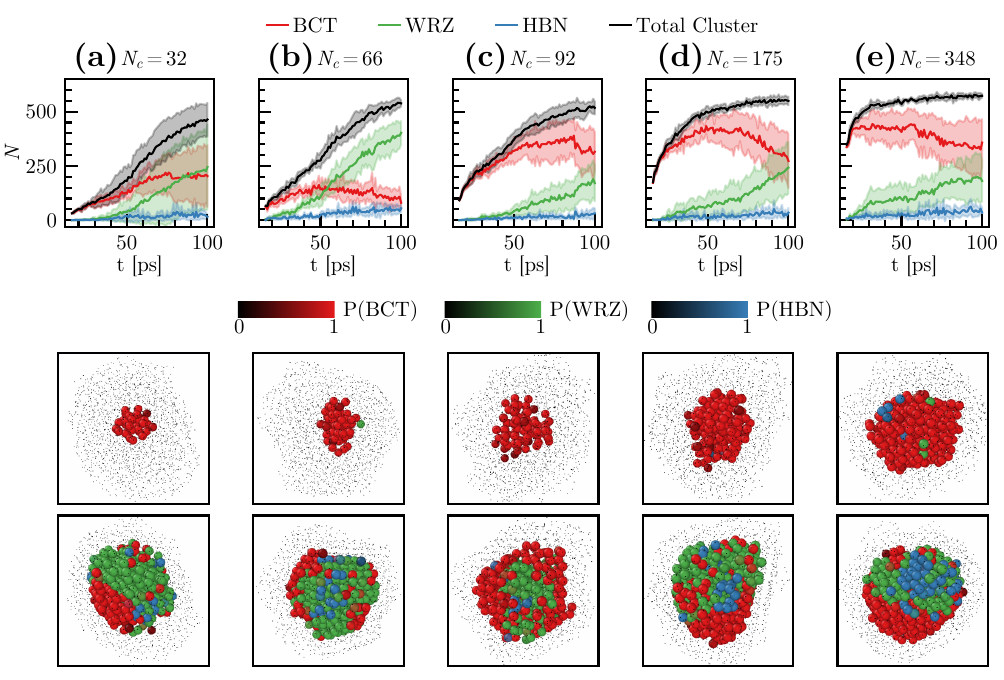}
    \caption{Extended growth simulations following the insertion of a  BCT crystal seeds at $1000$\,K for five different seed sizes in a 2000-atoms system. The composition of the crystalline clusters is shown in the graphs in each column. The first and last snapshots of each simulation are shown below the corresponding graphs. Atoms are colored according to their probabilities to be in BCT structure (red), WRZ structure (green), and HBN structure (blue).}
    \label{fig:longbct}
\end{figure*}

Then, the BCT case exhibits a much more complex picture. Indeed, from Fig.\,\ref{fig:longbct}, it can be seen that the composition of the crystalline cluster evolves with time. Right after the relaxation procedure, the crystalline clusters are composed almost entirely of atoms in the BCT structure as it is the structure introduced in the seeding initial phase. This means that our relaxation procedure correctly stabilizes the interface between crystal and liquid without changing the crystal structure. The number of atoms in the BCT phase then increases, but given enough time it starts to decline, and the number of atoms in the WRZ phase increases. Finally, at the end of the simulation, the center of the crystalline cluster is composed mostly of atoms in the WRZ structure, while the outer layers are composed of atoms in HBN and BCT structures. This observation suggests that the preferred crystallized structure for ZnO in these conditions remains WRZ even when starting with a BCT seed.

\section{Discussion}

The comparison between results from brute force and seeding simulations exhibits fundamental insights on the nucleation process. Indeed, at low temperatures associated with the deepest degree of supercooling, crystal nucleation seems to follow a multi-step process where a metastable phase (BCT) emerges first before being replaced by the most stable phase (WRZ). Meanwhile, at higher temperatures involving larger free energy barriers, the BCT seeds are in fact less stable than the WRZ ones. Indeed, we showed that their critical temperature is consistently lower than that of WRZ and that even when artificially inserted the BCT structure rapidly turns into WRZ during the subsequent growth. These two observations suggest that nucleation occurs in a single step for those moderate degrees of supercooling. To the best of our knowledge, such a change in nucleation pathway has so far only been found in crystal condensation from dilute systems like in the gas phase or in the presence of a surrounding liquid solvent. In the latter case, the observed metastable phase was made of a dense amorphous phase while here, it is an ordered phase made of different crystal polymorph. 

Despite the breadth of our findings, interpreting them remains a very challenging task. On the one hand, the preponderance of WRZ at moderate degrees of supercooling, ie. in the presence of free energy barrier, is consistent with a classical nucleation picture where the nucleating crystal is also the most stable crystal. On the other hand, it is more surprising that an alternative nucleation pathway involving a different crystal structure emerges at deeper degrees of supercooling when the free energy barrier is almost vanishing. As a possible explanation, a classical picture based on the capillary approximation would evoke the possibility that the crystal/liquid interface is more favorable for BCT than for WRZ which only becomes important under these temperature conditions since the nucleus is small enough to enhance surface effects. As such, it would be appealing to characterize bulk properties including crystal/liquid interface, migration rate or elastic stresses. While appealing for qualitative understanding, characterizing the crystal/liquid interface is rendered almost impossible because in this supercooled regime the interface is not stable and the crystal will spontaneously grow. Additionally, herein, the nucleus is composed of relatively few atoms (less than 50 atoms) and therefore it can not be characterized with any sort of bulk properties. More generally, the complexity herein is that even the concept of nucleus is no longer relevant in this regime of vanishing free energy barrier. As such, the current state of the art do not allow us to provide a quantitative explanation for our findings. However, it remains that from an empirical viewpoint, WRZ and BCT crystal symmetry can very well be compared structurally to face centered cubic (FCC) and body centered cubic (BCC). For the latter, the emergence of BCC when the free energy barrier vanishes has been observed in several occasions and can be explained using symmetry arguments and mean-field approaches\cite{Klein1986Dec}.



From the technical viewpoint, our work proposes three solutions to allow for studying nucleation in complex nanoscale systems. Firstly, we developed an approach for modeling long-range interactions that exploits LassoLars fitting for obtaining an effective static point charge Coulomb interaction. Our comparison with short-range PLIP shows that this approach enables us to better capture subtle charge effects related to polar surfaces which are crucial when dealing with nanoparticles where surface effects become preponderant. Secondly, we combined Gaussian Mixture Model with Steinhardt parameters to classify structures within a complex structural landscape made of 7 crystal polymorphs. We expect that the same methodology can be applied to characterize polymorphs in different materials as well as defects including grain boundaries or dislocations. Thirdly, we managed to explore nucleation at different degrees of supercooling by using brute-force simulations as well as the seeding technique. Altogether, our promising results advocate for transferring the proposed simulation strategy to the formation of different types of nanocrystals including quantum dots and nanoalloys.

\section*{Methods}

\subsection{Machine-learning interaction potentials with long-range physics}

\subsubsection{Short range PLIP}
To begin, we start with a concise description of the chosen short-range MLIP, before delving into long-range electrostatic interactions. To start, a linear model is employed to estimate each atomic energy ($E_{i}$) which is represented by a weighted sum of descriptors 
\begin{equation}
    E_{\textrm{short}}^{i} = \sum_{n}\omega_{n} \chi_n^{i}
\end{equation}
where the coefficients ($\omega_n$) are the fitting parameters. The descriptors for the PLIP model explicitly follow a many-body order expansion:
\begin{align}
&[\chi^{\textrm{2B}}]_n^i = \sum_j f_n(r_{ij}) \times f_{\textrm{c}}(r_{ij}), \\
&[\chi^{\textrm{3B}}]_{n,l}^i = \sum_j \sum_k f_n(r_{ij})f_{\textrm{c}}(r_{ij}) f_n(r_{ik})f_{\textrm{c}}(r_{ik})cos^l(\theta_{ijk}),  \\ 
&[\chi^{\textrm{NB}}]_{n,m}^i = \left( \sum_j f_n(r_{ij}) \times f_{\textrm{c}}(r_{ij}) \times f_{\textrm{s}}(r_{ij}) \right)^m,
\end{align}
where $r_{ij}$ is the distance between atoms $i$ and $j$, $\theta_{ijk}$ is the angle centered around the atom $i$, $f_n$ are a set of Gaussian functions with different widths and central positions, $f_{\textrm{s}}$ is a polynomial function that allows for setting the N-body interactions to $0$ at short range and $l$ and $m$ are two positive integers. The cutoff function is defined as:  $f_{\textrm{c}}(r_{ij}) = 0.5 \left(1 + {\textrm{cos}}(\pi(r_{ij}/r_{cut}))\right)$. In the following, we use $r_{cut}=6$~\AA  \,and impose that $l\leq 5$ and $m\leq 7$. For more details on the short-range PLIP model, one can refer to the following references\,\citep{Kandy2023May,Goniakowski2022Oct,Benoit2020Dec,Tallec2023Jan}.

\subsubsection{Electrostatic PLIP+Q}
The field of extending MLIP models to include long-range interactions is actively advancing, and recent developments in this area have been summarized by \citet{Anstine2023Mar}. In most of these approaches, a database of effective point charges is computed from electron structure calculations using different methods including Mulliken and Löwdin population analysis\,\cite{Mulliken1955Oct},  Bader analysis\,\cite{Bader1985Jan}, and the Density Derived Electrostatic and Chemical (DDEC) charge model\,\cite{Manz2016}. Then, a machine-learning model is constructed in order to determine on-the-fly the charge values based on the local environment surrounding each atom. More recently, further progress was obtained by training the long-range machine-learning model on susceptibility instead of charge values which enables for capturing subtle charge transfer mechanisms\,\cite{Ko2023Jun,Ko2021Feb,Ko2021Jan}. While these approaches remain the most accurate to date, they might suffer from implementation difficulty and computational costs. 

In our PLIP+Q model, we chose to use static point charges that are fixed with time and the local environment. In practice, we begin by setting the value of initial charges which can be for instance the oxidation number or deduced from electron density calculations. Then, we compute a fictive electrostatic contribution to the energies denoted $E^i_{\textrm{el}}$ using the point-charge Coulomb model along with the Ewald summation method\,\cite{ewald1921berechnung}. In order to determine how much this fictive electrostatic interaction indeed contributes to the overall interactions, we next define an overall charge scaling factor named $\gamma$ leading to the following total energy: 
\begin{equation}
\begin{aligned}
E^{\text{i}} &=  E_{\textrm{short}}^{i} + \gamma E_{\textrm{el}}^\text{i} 
\end{aligned}
\label{}
\end{equation}
As such, the scaling factor $\gamma$ becomes an additional linear coefficient that can be fitted along with those associated with the short-range interactions. We finally use the LassoLars regression algorithm to determine simultaneously the linear coefficients required for the short range interactions and the value of $\gamma$ thus enabling us to empirically obtain an effective value for the static point charge. Often in attempts at modeling long-range interactions using machine-learned charges, the short-range interactions is treated as a substantial difference between the long-range interactions and the total quantum accurate energy. Here, we chose instead to consider the long-range interactions simply as an additional descriptor without imposing its presence in the final model. As such, when using the LassoLars regression that selects only a subset of the most preponderant descriptors, the long-range interactions will be considered solely if necessary. In our present case, the initial charges are set as the oxidation number ie. at plus and minus 2 respectively for zinc and oxygen, and following the LassoLars fitting, the rescaled value of the charges becomes $\pm$0.64, to be compared with $\pm$ 1.16 obtained from the Bader decomposition of DFT charge densities. Although it is difficult to interpret this value because the short-range PLIP also contributes to the overall two-body interaction, it remains interesting to note that the LassoLars regression scheme is able to retrieve a value at a reasonable order of magnitudes.

\subsubsection{Training database}

The employed training database is exactly the same as in our two previous applications of PLIP to ZnO\,\cite{Kandy2023May,Goniakowski2022Oct} In brief, the reference DFT calculations are performed using the VASP software, employing the PW91 exchange-correlation functional and the projector augmented wave method. The standard zinc and soft oxygen pseudopotentials are employed, with an energy cutoff of 270 eV\, \cite{Kresse1993,Kresse1996,Kresse1999}. The total database is made of both ordered and disordered structures starting from bulk and surface structures.

In particular, six different ZnO polymorphs namely WRZ, zinc blend (ZBL), body-centered tetragonal (BCT), sodalite (SOD), h-BN (HBN), and cubane (CUB) crystallographic structures are included. We employ relatively large supercells in the range of 16-19\,\AA\,forming parallelepipeds that contain between 320 to 480 atoms. The Brillouin zone sampling is done at a single $\Gamma$ point. We conduct atomic coordinate relaxation until forces reduce below 0.01\,eV/\AA\,while maintaining stress tensor components below 0.01\,eV/\AA. To study nucleation in nanoparticles, it is also essential to include low-coordinated structures in the database. In this regard, for each of the six ZnO polymorphs, low-index nonpolar surfaces are also considered. In all our computations, we utilize slabs composed of 6 to 12 atomic layers, separated by approximately 15\,\AA\,of vacuum, and we conduct a thorough relaxation of atomic coordinates of all ions. 

From all of these equilibrium structures, we perform molecular dynamics (MD) simulations using either a classical potential or a previously obtained MLIP to sample additional structures where forces are then determined afterward with single-point DFT calculations. 

\begin{table}[ht]
\begin{tabular}{|l|r|}
\hline
\textbf{Shortname} & \textbf{Fullname}                 \\ \hline
BCT       & body-centered tetragonal \\ \hline
CUB       & cubane                   \\ \hline
HBN       & h-BN                     \\ \hline
SOD       & sodalite                 \\ \hline
WRZ       & wurtzite                 \\ \hline
ZBL       & zinc blend               \\ \hline
\end{tabular}
\label{CrystalStructure}
\caption{Nomenclature defining the acronyms used to describe each of the considered crystal structures.}
\end{table}

\subsection{Brute-force and seeding simulations}

\paragraph*{Brute-force simulations}

For the brute-force simulations, we used the NVT ensemble in the LAMMPS software\,\cite{plimpton1995fast} with a Nose-Hoover thermostat and a timestep of $1$\,fs. In order to work at a consistent degree of supercooling, we measure the melting temperature for a WRZ nanoparticle of each nanoparticle size using a linear temperature ramp [$2000$\,K.ns$^{-1}$] and fitting the crystal size with a hyperbolic tangent function. The melting temperatures are $1213$\,K, $1401$\,K, $1503$\,K, and $1531$\,K for sizes of $500$, $1000$, $1500$, and $2000$ atoms, respectively. We note that our MD simulations as well as experimental measurements seems to indicate that the ZnO melting is congruent\cite{Mukhanov2013Jul}. Despite imposing the same degree of supercooling, the nucleation induction time slightly differs for each nanoparticle and we used different simulation duration, between $1$ and $10$\,ns, as necessary for the different droplet sizes. By varying initial atomic velocities we obtained multiple simulation trajectories for the same system and temperature. By doing this we are able to assess the average behavior of this type of system at the chosen temperature.

\paragraph*{Seeding simulations}

After initializing the system with a crystalline seed of a chosen size $N_0$ inserted inside a liquid nano-droplet, the relaxation protocol enabling for reaching the desired temperature $T$ consists of three different steps. Firstly, the crystalline cluster is kept static while MD is performed on the liquid nano-droplet in the NVT ensemble at the relaxation temperature $T_0$, which was chosen always lower than the expected critical temperature and changed depending on the size of the droplet and the crystalline cluster. This step is performed for a duration that is adjusted according to the system size in order to avoid complete crystallization during this step. Because only the liquid droplet can move, the size of the crystal seed can increase through interfacial growth. Furthermore, the gap created during the insertion of the seed between the crystal and the liquid is filled during this first step. Secondly, atoms in the crystalline cluster are also allowed to move in the NVT ensemble at an increasing temperature. In this second step, of the same duration as the first one, two separate Nose-Hoover thermostats are used: (1) for the liquid droplet at $T_0$ and (2) for the crystalline seed going from $100$\,K to $T_0$. During this step, we attempt to keep the size of the cluster constant. The purpose of this step is to slowly increase the temperature of the crystalline cluster until $T_0$ is achieved for the whole system. Thirdly, the temperature of the system needs to go from $T_0$ to $T$. For that purpose, we can not directly set the thermostat temperature at $T$ as it would put the system out of equilibrium. In addition, because nucleation is a stochastic process, it was crucial to be able to study the same system (ie. size of crystalline seed + temperature) with different initial velocities. For these reasons, the Langevin thermostat\,\cite{ref:langevin} is used at an increasing temperature from $T_0$ to $T$ and with $5$ different random values for the thermostat. At the end of these three steps, the apparent size of the crystalline seed that is used for the analysis is always different from $N_0$ and is denoted $N_c$. After the Langevin temperature ramp, NVT simulations using the traditional Nose-Hoover thermostat are carried out at a temperature of interest $T$. At the end, the obtained growth/melting curves are averaged over the different random values of the Langevin thermostat thus accounting for the stochasticity of the nucleation process. We used those 5 different simulations to also compute a standard deviation associated to each growth/melting curves. The duration of each step and the relaxation temperatures used are presented in Supplementary Table 1.



\subsection{Gaussian-mixture model to characterize polymorphic crystal ordering}

\begin{figure*}[ht]
    \centering
    \includegraphics[width=17cm]{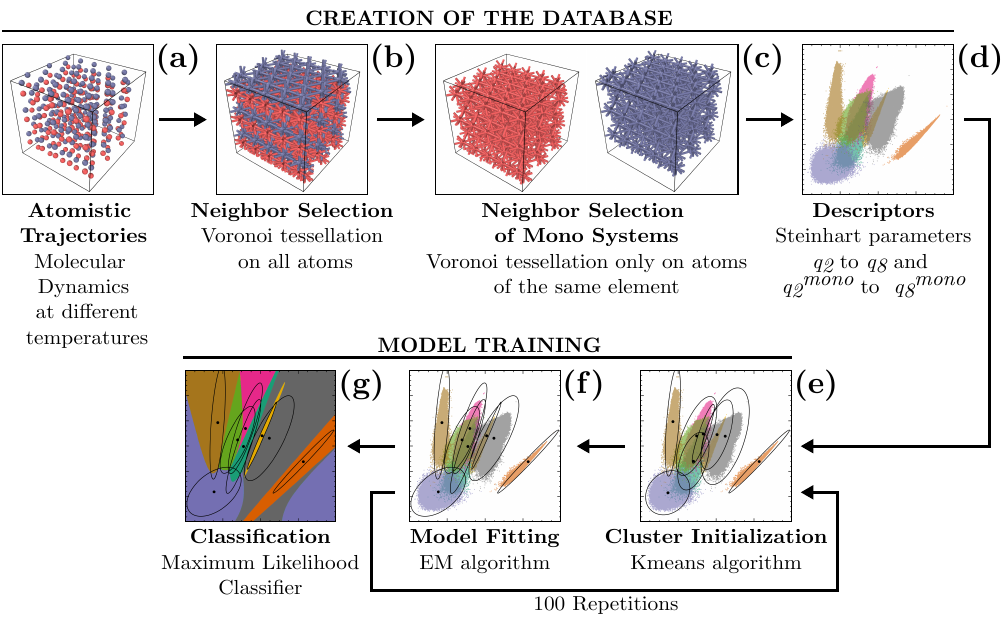}
    \caption{Schematic of the steps in SGMA. A database is created from (a) atomistic trajectories obtained using Molecular Dynamics. (b) The neighbors of each atom are found using Voronoi tessellation, (c) as well as the neighbors of only the same element. (d) The Steinhardt parameters are then computed for all atoms in each snapshot of the trajectory. A Gaussian Mixture Model is then trained on the database. (e) The parameters of the Gaussian clusters are initialized using the Kmeans algorithm and are then optimized using the Expectation-Maximization algorithm (f). Steps (e) and (f) are performed 100 times and the parameters with the best results are kept. (g) Classification is then performed on the structures in the database or on new test structures.}
    \label{fig:structuralAnalysis}
\end{figure*}

In order to analyze local ordering in the obtained simulations, it is crucial to use a numerical tool capable of distinguishing 7 different polymorphs of ZnO among which 6 are already employed in the DFT training database as well as the rock salt (RCK) structure that was considered only for the structure identification. We propose a supervised learning method that we call Steinhardt Gaussian Mixture Analysis (SGMA) [See Fig.\,\ref{fig:structuralAnalysis} for a schematic picture]. 

In particular, we start by creating a database consisting of crystalline structures sampled around their equilibrium positions. For that purpose, NVT simulations are performed during 10\,ps with an increasing temperature from $200$\,K to $1500$\,K controlled via a Nosé-Hoover thermostat. The duration and the upper temperature are chosen so that none of the seven considered crystals melt. In addition, liquid structures are also sampled in our database using NVT simulation at $2500$\,K. 21 snapshots are extracted along those simulations for each of the crystal polymorphs and for the liquid. 

For each of these snapshots, we then compute the averaged Voronoi weighted Steinhardt parameters\,\cite{ref:MickelKapfer} using the \textit{Pyscal}\,\cite{Menon2019Nov} library in \textit{Python}. We augmented the generic Steinhardt values with homo nuclear ones calculated after removing each hetero atom type. In both cases, we used Steinhardt parameters indexed from 2 to 8  thus making a list of 14 order parameters to characterize the local ordering of each atom in a given snapshot.

The training of the database is next performed using the Gaussian Mixture Model (GMM) as implemented in the \textit{scikit-learn}\,\cite{Pedregosa2011} library in \textit{Python}. The unknown parameters of the GMM were iteratively estimated using the Expectation-Maximization algorithm\,\cite{ref:EM}. The GMM was trained using full covariance matrices and 100 k-means initializations. In our case, instead of letting the GMM determine the number of Gaussian components automatically, we chose to impose it equal to the number of structure types in our database ie. 8. In this way, we give priority to the physical meaning of our database. For classification, the Maximum Likelihood Classifier is utilized, in which the probability of an object $x_i$ to belong to class $\omega_k$ is computed as: 
$    p(\omega_k | \mathbf{x}_i) = \frac{\alpha_k \mathcal{N} (\mathbf{x}_i | \mathbf{m}_k, \mathbf{C}_k)}{\sum_{j = 1}^{K} \alpha_j \mathcal{N} (\mathbf{x}_i | \mathbf{m}_j, \mathbf{C}_j)},$ where $\alpha_k$ are mixture proportions and $\mathbf{m}_k$ and $\mathbf{C}_k$ are the mean vector and the covariance matrix of each Gaussian component $\omega_k$. The mixture proportions satisfy the conditions $0 \leq \alpha_k \leq 1$ and $\sum_{k = 1}^{K} \alpha_k = 1$. These values can then be interpreted as the probability of an atom being in one of the 8 structure types in the database. In the Maximum Likelihood Classifier, the probabilities are usually compared and an object is said to belong to a category for which it has the highest probability. In this work, we chose a more severe classification rule and considered an atom to belong to a cluster only when its probability is higher than 50\%.

Parameters for the different Gaussian clusters in the model are obtained following the training procedure. However, only the index of each cluster is known and no information is given as to what structure each cluster represents. To do this, the model is tested by predicting the Gaussian cluster to which the perfect crystals and the liquid belong. In this way, the labels of each Gaussian cluster are obtained. This is another way in which our method differs from previous uses of the GMM. We first train a model, and since the clusters in our database are approximately Gaussian, it is expected that after training a Gaussian cluster will be assigned to each structure type. This model can then be used to analyze systems different from the ones encountered in the database and obtain specific structural predictions in a supervised manner.

Altogether, our SGMA methodology allows us to predict crystal structures in a system by creating a database with the different known polymorphs. In this application, we rely on the assumption that each structure in our database can be approximated by a single Gaussian cluster, as opposed to other applications of the GMM where the number of Gaussian clusters is found automatically\,\cite{ref:beckerDevijver, ref:BoattiniAguilar, ref:BoattiniDijkstra}. For future applications, the method can also be adapted to automatically find the number of required Gaussian clusters when a structure is better represented with more than one. Our method is also characterized by the significantly large number of descriptors that are computed and used for analysis. Compared to other methods, we do not make use of dimensionality reduction techniques to decrease the complexity of the data at the clustering step\,\cite{ref:BoattiniAguilar, ref:BoattiniDijkstra, Coslovich2022Nov, ref:gasparottoMeissner, Goniakowski2022Oct, Pipolo2017Dec, ref:Reinhart, Sarupria2022Sep, ref:TamuraMatsuda}. If the computational cost demands it, it is possible to reduce the complexity of the model by carefully choosing the descriptors that distinguish the structures in the database the best. 


\begin{figure}[ht]
\includegraphics[width=8.6cm]{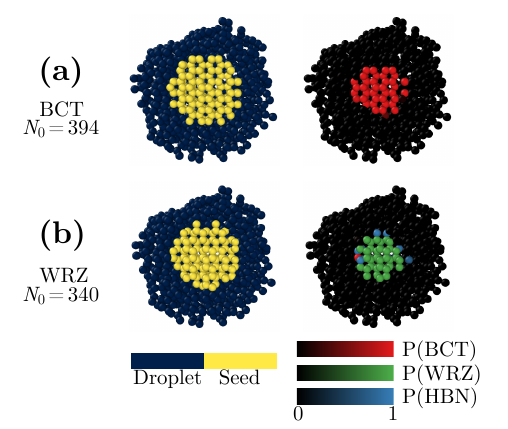}
\caption{Test of our structural analysis methodology on BCT (a) and WRZ (b) seeded nanoparticles. The left column illustrates the crystalline atoms that were inserted in the droplet system. The right column shows the predicted structure of each atom as computed with our methodology.}
\label{fig:testgmm}
\end{figure}

To test the usage of SGMA in the seeding technique we analyzed the structure of seeded nanoparticles right after the initialization and before the relaxation step for BCT and WRZ crystal seeds, illustrated in Fig.\,\ref{fig:testgmm}. It was found that in both cases our model correctly identifies the crystalline cluster surrounded by the liquid. In the case of the BCT crystal seed, the number of atoms predicted to be in the BCT phase is lower than the number of inserted atoms. This is expected from our model since at the interface between crystal and liquid, the atomic environments differ significantly from the environment represented in the database. Even though the atoms at the interface still present some order in their structure, according to the Maximum Likelihood Classifier described previously, they are considered to be closer to the liquid structure. Similarly, in the case of the WRZ crystal seed, the number of atoms predicted to be in the crystalline phase is lower than the number of inserted atoms. This time, however, not all atoms in the crystalline cluster are predicted to be in the WRZ phase. At the surface of the crystalline cluster, some atoms are predicted to be in the HBN or BCT phases instead. This is due to the large resemblance between WRZ and HBN phases. With these tests, we have shown the physical meaning of the predictions performed using the GMM. Crystalline atoms were correctly identified when surrounded by a liquid, with the exception of atoms at the interface between liquid and WRZ phases that can be also labeled as HBN. These results confirm that the training parameters of the model were appropriately chosen, and the physical meaning of the database was retained.

\section{Data availability} 
The authors declare that the data supporting the findings of this study are available within the article and its supplementary information files or from the corresponding authors on reasonable request.

\section{Code availability} 
The implementation of (1) PLIP+Q, (2) the structural analysis, (3) Seeding and brute forces simulations inputs and outputs will be shared with community upon request.


\section{Acknowledgments}

This study was supported by the French National Research Agency (ANR) in the framework of its “Jeunes chercheuses et jeunes chercheurs” program, ANR-21-CE09-0006. This work was performed using HPC/AI resources from GENCI-[IDRIS/TGCC] (Grant 2021,2022-A0110913010) and using CALMIP (Grant 2021,2022-P21004).

\section{Author contributions}
C. R. S. performed and analysed the molecular dynamics simulations and developed the approach for structural analysis. A. K. A. K. established the PLIP+Q method and performed the comparison with PLIP. Both C. R. S. and A. K. A. K. contributed equally to the work. J. F. participated in the development of the approach for structural analysis. Q. M. participated in the comparison between PLIP and PLIP+Q. J. G. performed the DFT calculations that were required to establish the database along with the comparison with DFT. J. L. supervised the work and funded all parts of the project. All authors contributed to the writing and reviewing of the manuscript.              

\section{Competing interests}
The authors declare no competing interests.

\end{document}